\documentclass[twocolumn,twoside,slac_two]{revtex4}
\usepackage{graphicx}
\usepackage{fancyhdr}
\usepackage{subfigure}
\begin{document}
\voffset=-1cm
\title{Annihilation of light dark matter into photons in model-independent approach}

%

\author{Andriy Badin\footnote{presenting author} \vspace{5pt}}
\email{a_badin@wayne.edu}
\affiliation{Department of Physics and Astronomy\\
        Wayne State University, Detroit, MI 48201}\vspace{10pt}
\author{\vspace{10pt}Alexey A.\ Petrov\vspace{5pt}}
\email{apetrov@wayne.edu}
\affiliation{Department of Physics and Astronomy\\
        Wayne State University, Detroit, MI 48201}
\affiliation{\vspace{5pt} Michigan Center for Theoretical Physics\\
University of Michigan, Ann Arbor, MI 48109\vspace{10pt} \\
$\phantom{}$}
\author{Gagik K. Yeghiyan\vspace{5pt}}
\email{ye_gagik@wayne.edu}
\affiliation{Department of Physics and Astronomy\\
        Wayne State University, Detroit, MI 48201}
\begin{abstract}
We examine annihilation of light bosonic Dark Matter into pair of
photons in model-independent way. We consider the simplest generic
Lagrangian describing such process and then compare results to the
available experimental data.  Then we match our results with
particular Dark matter models and determine possible constrains onto
parameter space of those models.
\end{abstract}

\maketitle

\thispagestyle{fancy}



The presence of cold Dark Matter (DM) in the Universe provides
explanations to several observational puzzles and is an established
fact nowadays. However despite numerous experimental efforts the
nature of DM remains a mystery. Many elementary particle theories
beyond the Standard Model (SM) have in their content at least one
electrically neutral, stable, weakly interacting particle. In order
to make selection between those models a consideration of DM
properties using different observables with the minimal number of
assumptions is needed. In this paper we provide a model-independent
approach for annihilation of light cold scalar dark matter. Such
limitations are motivated by WMAP observations which ruled out warm
DM \cite{WMAP} and the fact that Lee-Weinberg limit that forbids
light dark matter can be avoided for non-fermionic Dark Matter
particles \cite{Boehm}. It means that independent constraints of
such models are useful tool for discrimination of different
theories. It can be argued that combined constraints from heavy
quarkonium decays, astrophysical observation, and direct DM
detection experiments can limit parameter space of such Dark Matter
candidates \cite{us}. Here we report on astrophysical observations.

Due to large available amount of data we used gamma ray flux as an
experimental observable to constrain properties of DM . We compare
data from EGRET with theoretical calculations of flux from
$\phi\phi\rightarrow\gamma\gamma$ process. This process is
suppressed compared to the $\phi\phi\rightarrow X\gamma$, however it
provides very distinct spectrum feature and is very easy detectable.
Current data from EGRET \cite{EGRET,EGRET1} do not have any signs of
monochromatic lines in photon spectrum, meaning that flux from
$\phi\phi\rightarrow\gamma\gamma$ is below diffuse background. This
fact can be used to derive constraints on properties of Dark Matter.

The paper is organized in the following way. We introduce generic
model-independent lagrangian describing DM-photon and compute gamma
ray flux from DM annihilation.

\section{Photon flux from DM annihilation}

In general, annihilation process can be described with an effective
Lagrangian of the following form:
\begin{equation}
\label{photon_general} {\cal L}_{eff}=A_1\phi\phi
F_{\mu\nu}F^{\mu\nu}+A_2\phi\phi \tilde{F}_{\mu\nu}F^{\mu\nu}
\end{equation}
Any additional operators will be of higher dimension and therefore
their contribution will be suppressed. Completing the textbook level
calculation of the cross-section one can get
\begin{eqnarray}
\sigma= \frac {s^2}{8\pi \sqrt{s(s-4\mu^2)}} (|A_1|^2 + 2 |A_2|^2)
\end{eqnarray}

 Assuming that we deal with non-relativistic light dark matter, we
present transferred energy as
\begin{equation}
s \approx
4\mu^2+(\vec{p_1}-\vec{p_2})^2=(2\mu)^2(1+(\frac{\vec{v_1}-\vec{v_2}}{2})^2),
\end{equation}
\flushleft where $\mu$ is a mass of Dark Matter particle.
 Another assumption we make is that DM particles are
distributed according to Maxwell-Boltzman distribution. After we
expand cross-section around $s=4\mu^2$ and average it with MB
distribution
\begin{eqnarray}
\sigma v &=& \frac{s^{3/2}(|A_1|^2+2|A_2|^2)}{8\pi\mu}\approx A+B(\vec{v_1}-\vec{v_2})^2\\
\langle\sigma v\rangle &=& A+B\frac{6kT}{\mu}
\end{eqnarray}
where $A$ and $B$ denote the following combinations of Wilson
coefficients $A_i$
\begin{eqnarray}
A&=&\mu^2(|A_1|^2+2|A_2|^2)\\
B&=&\mu^2(\frac{3}{8}(|A_1|^2+2|A_2|^2)+\\
&+&\mu^2(2Re[A_1\frac{\partial A_1}{\partial
s}]+4Re[A_2\frac{\partial A_2}{\partial s}]))\nonumber,
\end{eqnarray}
\flushleft with their values taken at $s=4\mu^2$. It is worth
pointing out that a fraction $\frac{6kT}{\mu}\sim\frac{v^2}{c^2}\ll
1$ which means that in most cases the contribution from the second
term is negligible.

The differential flux of photons produced by DM annihilations is
\cite{remark}
\begin{equation}
I(E,\psi)=\frac{dN_{\gamma}}{dE}\frac{\langle\sigma
v\rangle}{2\mu^2}J(\psi)
\end{equation}
where $\psi$ is  the angle between the galactic center and the line
of observation, and
\begin{equation}
J(\psi)=\int_{l.o.s}ds\frac{\rho^2[r(s,\psi)]}{4\pi}
\end{equation}
is an integral along line of sight which depends on the choice of
dark mater halo profile. As it was argued in \cite{Bergstrom}, the
maximum flux will be in the direction of the galactic center. The
highest value of flux is given by the choice of Navarro-Frenk-White
(NFW) profile for the DM distribution. This will provide us with the
upper limit on the theoretical value of the photon flux and on the
parameters of dark matter. Using results from the same paper we
obtain:
\begin{equation}
I(E,\psi)=7.3\times10^{-5}\frac{dN_{\gamma}}{dE}\frac{\langle\sigma
v\rangle}{2\mu^2} cm^{-1} s^{-1} sr^{-1} GeV^{-1}
\end{equation}
 In this result the dependence on the particle physics dynamics
is separated from the structure of Dark Matter halo.

\begin{figure}
\centering
\subfigure[\ Direct channel] 
{
    \label{fig:sub:a}
    \includegraphics[width=4cm]{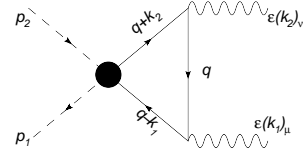}
} \hspace{1cm}
\subfigure[\ and a crossed one ] 
{
    \label{fig:sub:b}
    \includegraphics[width=4cm]{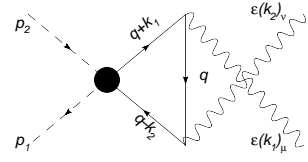}
} \caption{Diagrams contributing to DM annihilation}
\label{fig:sub} 
\end{figure}

To proceed further we need to introduce mechanism of DM
annihilation. As electrically neutral field DM can not be coupled to
the photons directly. It is natural to assume that for the light
Dark Matter ($\mu < 5 GeV$) the only relevant couplings are the ones
that couple it to the Standard Model fermions. If we limit ourselves
to the operators of the highest possible dimension six, the
effective Lagrangian will take the form

\begin{eqnarray}
\label{effecitve_lagrangian} -{\cal L}&=&\frac{2}{\Lambda^2}(C_1
O_1+ C_2 O_2)
\end{eqnarray}
where $\Lambda$ is a heavy mass scale, for example mass of heavy
mediator that provides the interaction between the SM and the DM
sectors. The operators are defined as
\begin{eqnarray}
O_1&=& m_f \phi\phi\bar{\psi}\psi\\
O_2&=&i m_f \phi\phi\bar{\psi}\gamma_5\psi
\end{eqnarray}
The choice is such that they are hermitian and their Wilson
coefficients $C_i$ are real. $\psi$ are the SM fermion fields. There
are only two types of diagrams that contribute to the annihilation
process ( Fig.\ref{fig:sub} )

For computation of annihilation rate the contribution from all
possible fermions should be taken into account (i.e. summation over
leptons and quarks performed). Summation is assumed and charge of
loop fermions is denoted as $Q_f$ in the analytical expressions
presented further in text. Explicit calculation of those diagrams
using introduced generic lagrangian leads to the following Wilson
coefficients $A_{1,2}$:
 \begin{eqnarray}
 A_1&=&\sum_f\frac{16 C_1 m_f^2 \pi^2
 Q_f^2}{s \Lambda^2}((4m_f^2-s)C_0(0,0,s,m_f^2,m_f^2,m_f^2)+2)\nonumber\\
A_2&=&-\frac{16}{\Lambda^2} \sum_f\imath C_2 m_f^2 \pi^2 Q_f^2
C_0(0,0,s,m_f^2,m_f^2,m_f^2)
\end{eqnarray}

Where, $s$ is the Mandelstam variable $$s=(p_1+p_2)^2=(k_1+k_2)^2$$
and
\begin{eqnarray}
&&C_0(p_1^2,
(p_1-p_2)^2,p_2^2,m_1^2,m_2^2,m_3^2)=\nonumber\\
&=&\int\frac{d^4q}{\imath
\pi^2}\frac{1}{(q^2-m_1^2)((q+p_1)^2-m_2^2)
((q+p_2)^2-m_3^2)}\nonumber
\end{eqnarray}
is a Passarino-Veltman three-point function, and for particular set
of parameters arising here it can be expressed analytically in the
following form:
\begin{equation}
C_0(0,0,s,m^2,m^2,m^2)=-\frac{1}{s}\tan^{-1}(\frac{\sqrt{s}}{\sqrt{4m^2-s}})^2
\end{equation}
It is worth pointing out, that our result for annihilation cross
section via channel governed by operator $O_1$ essentially
reproduces the result of \cite{Shifman} for annihilation of Higgs
boson into two photons and does not vanish in the  heavy fermion
mass limit. Also, when the mass of the dark matter particle is close
to the mass of the fermion in the loop, some low energy resonance
states that increase annihilation cross-section might appear. Such a
situation needs special treatment and is not considered here.

Let us now use the  experimental data to put constraints onto $C_1$
and $C_2$. Experimental data from EGRET can be parameterized in the
following way \cite{EGRET,EGRET1}:
\begin{eqnarray}
I&=&I_{gal} + I_{ex}\  cm^{-2}s^{-2} sr^{-2}GeV^{-1}\nonumber\\
I_{ex}&=&(7.32\pm0.34)\times 10^{-6}\left(\frac{E}{0.451
GeV}\right)^{-2.10\pm0.03}\\
I_{gal}&=&N_0(l,b)\times 10^{-6}
\left(\frac{E}{GeV}\right)^{-2.7}\nonumber
\end{eqnarray}
where
\begin{eqnarray}
&&-180^\circ\leq l \leq 180^\circ\mbox{ and } -90^\circ\leq b \leq 90^\circ\nonumber\\
N_0(l,b)&=&0.5+\frac{85.5}{\sqrt{1+(l/35)^2}\sqrt{1+(b/1.8)^2}}\\
&&\mbox{\ for \ }|l|\leq30\nonumber\\
N_0(l,b)&=&0.5+\frac{85.5}{\sqrt{1+(l/35)^2}\sqrt{1+[b/(1.1+0.022|l|)]^2}}\nonumber\\
&&\mbox{\ for \ }|l|\geq30\nonumber
\end{eqnarray}

\begin{figure}
\centering
\subfigure[\ Static DM] 
{
    \label{fig:limits:a}
    \includegraphics[width=6cm]{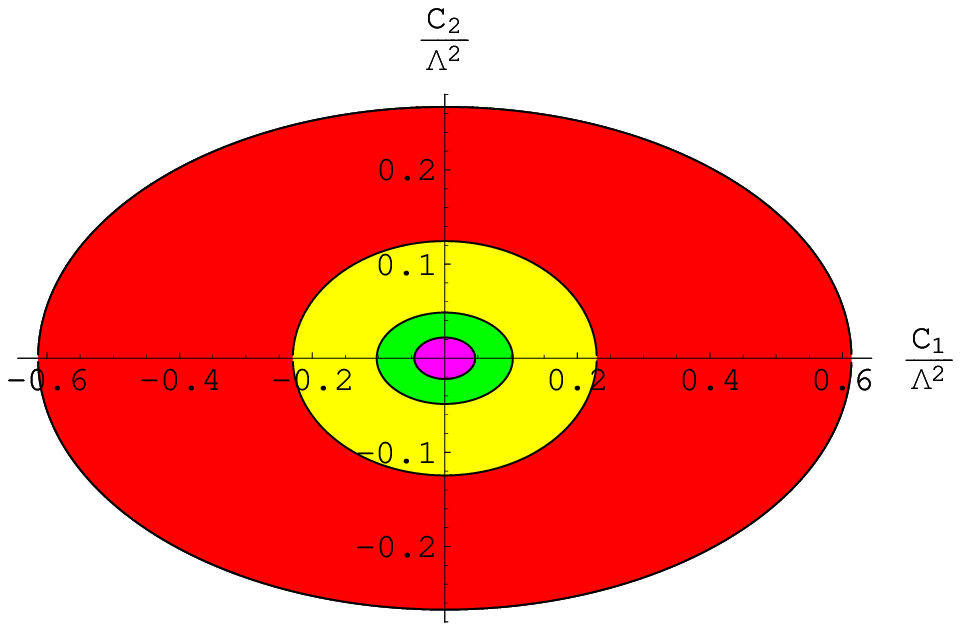}
}
\subfigure[\ Relative motion taken into account] 
{
    \label{fig:limits:b}
    \includegraphics[width=6cm]{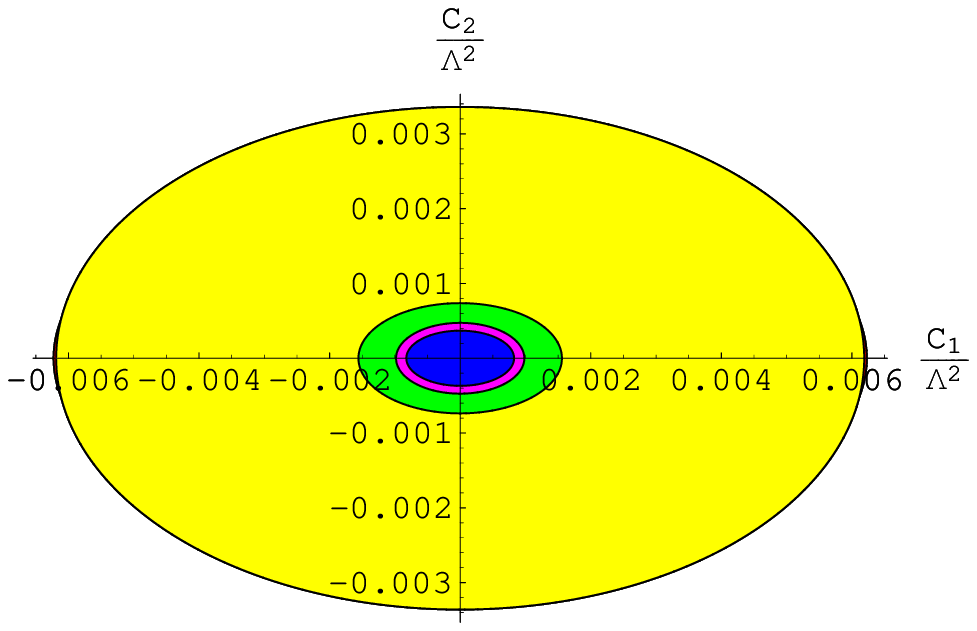}
} \caption{Constrains on DM parameter space in case of static (a)
and moving (b) DM. Filled regions are allowed parameter space for DM
particles of different mass: red -$\mu=0.1GeV$, green -$\mu=0.5GeV$,
blue -$\mu=1GeV$, pink -$\mu=2GeV$ and yellow - $\mu=5GeV$}
\label{fig:limits} 
\end{figure}

Since the highest flux will be from the direction of the galactic
center, we need to compute flux at $l,b=0$. There was no
monochromatic peak observed at EGRET, which means that intensity of
flux from dark matter annihilation is less than diffuse background.
This leads to the following constraining condition:
\begin{equation}
\frac{I_{theory}}{UpperBound(I_{ex}+I_{gal})}\leq1
\end{equation}
Assuming annihilation of DM particles that are at rest, photon
spectrum will be a monochromatic line with $E_{\gamma}=\mu$.
Detector measuring spectrum has finite resolution, thus instead of
$\delta$-function a spectrum integrated over some region of energies
will be measured. Considering several different masses of DM
particles leads to the following constrains that can be placed on
coupling constants :
\begin{eqnarray}
\label{limits1}
&&2.657\left(\frac{C_1}{\Lambda^2}\right)^2+14.02\left(\frac{C_2}{\Lambda^2}\right)^2\leq1\nonumber\\
&&\mbox{\
for \ }\mu=0.1 GeV\nonumber\\
&&94.89\left(\frac{C_1}{\Lambda^2}\right)^2+424.1\left(\frac{C_2}{\Lambda^2}\right)^2\leq1\nonumber\\
&&\mbox{\
for \ }\mu=0.5 GeV\nonumber\\
&&479.0\left(\frac{C_1}{\Lambda^2}\right)^2+2396\left(\frac{C_2}{\Lambda^2}\right)^2\leq1\nonumber
\mbox{\
for \ }\mu=1.0 GeV\\
&&477.6\left(\frac{C_1}{\Lambda^2}\right)^2+2057\left(\frac{C_2}{\Lambda^2}\right)^2\leq1\nonumber\\
&&\mbox{\
for \ }\mu=2.0 GeV\nonumber\\
&&19.02\left(\frac{C_1}{\Lambda^2}\right)^2+64.55\left(\frac{C_2}{\Lambda^2}\right)^2\leq1\nonumber\\
&&\mbox{\ for \ }\mu=5.0 GeV\nonumber
\end{eqnarray}
which are presented graphically on the Fig.\ref{fig:limits:a}

However, in real-life situation the spectrum will be smeared due to
thermal motion of DM particle, orbital motion of Earth, etc. and
therefore energy spectrum will have a shape of  a peak of finite
width and finite height instead of $\delta$-function. Assuming good
enough resolution of experimental set up, this peak might be
detected. Thermal velocity of DM particles is taken to be $v_{DM}=9
km/s$ \cite{DM_speed} and taking into account orbital motion of
Earth we approximate spectrum as a Gaussian distribution with
$\sigma=\mu v/c$ where $v=v_{DM}+v_{orbital}$. Such a choice of
$v_{DM}$ among all experimental data provides us with the most
narrow and high peak. It will be very easy detectable and will lead
to the highest values of upper bounds on DM model parameters. The
results after relative motion is considered are presented at
Eq.\ref{limits2} and Fig.\ref{fig:limits:b}
\begin{eqnarray}
\label{limits2}
&&2.58\times10^4\left(\frac{C_1}{\Lambda^2}\right)^2+1.36\times10^5\left(\frac{C_2}{\Lambda^2}\right)^2\leq1\nonumber\\
&&\mbox{\ for \ }\mu=0.1GeV\nonumber\\
&&4.12\times10^5\left(\frac{C_1}{\Lambda^2}\right)^2+1.84\times10^6\left(\frac{C_2}{\Lambda^2}\right)^2\leq1\nonumber\\
&&\mbox{\ for \ }\mu=0.5GeV\nonumber\\
&&1.47\times10^{6}\left(\frac{C_1}{\Lambda^2}\right)^2+7.35\times10^6\left(\frac{C_2}{\Lambda^2}\right)^2\leq1\nonumber\\
&&\mbox{\ for \ }\mu=1.0GeV\\
&&1.04\times10^{6}\left(\frac{C_1}{\Lambda^2}\right)^2+4.46\times10^6\left(\frac{C_2}{\Lambda^2}\right)^2\leq1\nonumber\\
&&\mbox{\ for \ }\mu=2.0GeV\nonumber\\
&&2.61\times10^4\left(\frac{C_1}{\Lambda^2}\right)^2+8.86\times10^4\left(\frac{C_2}{\Lambda^2}\right)^2\leq1\nonumber\\
&&\mbox{\ for \ }\mu=5.0GeV\nonumber
\end{eqnarray}

Realistically, after taking into account all possible effects,
contsrains will be somewhere between  ones provided in
Eq.\ref{limits1}, Fig.\ref{fig:limits:a} and ones given in
Eq.\ref{limits2}, Fig.\ref{fig:limits:b}

\section{Model of SM singlet scalar DM as an example}
As an example we consider DM annihilation in framework of Minimal
Scalar Dark Matter model (see for example \cite{Pospelov}). In this
model DM interaction with Standard Model fields is mediated by
exchange of a Higgs boson. The model is restricted based on relic
abundance calculations, however due to its simplicity it is perfect
for testing of our approach. The matching conditions for the Wilson
coefficients have the following form:
\begin{eqnarray}
\label{ScalarDM}
C_1&=&\lambda/2\nonumber\\
C_2&=&0\\
\Lambda&=&M_h\mbox{\ with\ }M_h\geq115GeV\nonumber
\end{eqnarray}
Inserting these parameters into model-independent bounds derived in
Eq.\ref{limits1} and Eq.\ref{limits2} leads to the following
constrains onto parameters of this model:
\begin{eqnarray}
\label{Scalar_DM_limits1}
|\lambda|&\leq&16227\left(\frac{M_h}{115}\right)^2\mbox{\ for \ }\mu=0.1GeV\nonumber\\
|\lambda|&\leq&2715.3\left(\frac{M_h}{115}\right)^2\mbox{\ for \ }\mu=0.5GeV\nonumber\\
|\lambda|&\leq&1208.5\left(\frac{M_h}{115}\right)^2\mbox{\ for \ }\mu=1.0GeV\\
|\lambda|&\leq&1210.4\left(\frac{M_h}{115}\right)^2\mbox{\ for \ }\mu=2.0GeV\nonumber\\
|\lambda|&\leq&6064\left(\frac{M_h}{115}\right)^2\mbox{\ for\
}\mu=5.0GeV\nonumber
\end{eqnarray}
for case of static DM and
\begin{eqnarray}
\label{Scalar_DM_limits2}
|\lambda|&\leq&164\left(\frac{M_h}{115}\right)^2\mbox{\ for \ }\mu=0.1GeV\nonumber\\
|\lambda|&\leq&41.2\left(\frac{M_h}{115}\right)^2\mbox{\ for \ }\mu=0.5GeV\nonumber\\
|\lambda|&\leq&21.8\left(\frac{M_h}{115}\right)^2\mbox{\ for \ }\mu=1.0GeV\\
|\lambda|&\leq&26.0\left(\frac{M_h}{115}\right)^2\mbox{\ for \ }\mu=2.0GeV\nonumber\\
|\lambda|&\leq&163.7\left(\frac{M_h}{115}\right)^2\mbox{\ for\
}\mu=5.0GeV\nonumber
\end{eqnarray}
if we take thermal motion of Dark Matter particles into
consideration.

As one can see, the obtained constraints are not very restrictive
for this particular model. However, consideration of the models with
enhanced couplings (for example two Higgs doublet model) provides
more strict constraints onto the parameters of the model \cite{us}.


\begin{thebibliography}{99}
\bibitem{remark} We consider self conjugated particles. For case of
non-self conjugated DM extra factor of $1/2$ is needed.
\end{thebibliography}
\end{document}